\documentclass{elsart}
\usepackage{amsmath,graphicx,amssymb}
\newcommand{\uE}{\ensuremath{\mathrm{E}}}  
\newcommand{\be}{\ensuremath{\frac{\mathrm{d} B (\mathrm{E}1)}
{\mathrm{d}E}}}                            
\newcommand{\urms}{\ensuremath{\mathrm{rms}}}
\newcommand{\ucm}{\ensuremath{\mathrm{cm}}}
\newcommand{\uc}{\ensuremath{\mathrm{c}}}
\begin{document}
\begin{frontmatter}
\title{Analytical E1 strength functions of
two--neutron halo nuclei: the \nuc{6}{He} example}
\author[chalmers]{C.~Forss\'en\corauthref{cor}},
\corauth[cor]{Corresponding author.}
\ead{c.forssen@fy.chalmers.se}
\author[kurchatov]{V.~D.~Efros} and
\author[chalmers]{M.~V.~Zhukov}
\address[chalmers]{Department of Physics, Chalmers University of
Technology and G{\"o}teborg University, S--412~96 G{\"o}teborg, Sweden}
\address[kurchatov]{The Kurchatov Institute, 123182 Moscow, Russia}
\begin{abstract}
An analytical model is developed to study the spectra of electromagnetic
dissociation of two--neutron halo nuclei without precise knowledge about
initial and final states. Phenomenological three--cluster bound state
wave functions, reproducing the most relevant features of these nuclei,
are used along with no interaction final states. The \nuc{6}{He}
nucleus is considered as a test case, and a good agreement with
experimental data concerning the shape of the spectrum and the magnitude
of the strength function is found.
\end{abstract}
\begin{keyword}
Borromean halo nuclei, three--body model, electromagnetic dissociation,
strength function, continuum excitations
\PACS 25.70.De; 21.60.Gx; 24.10.-i; 27.20.+n
\end{keyword}
\end{frontmatter}
\section{Introduction}
Coulomb excitation reactions serve as one of the most powerful tools for
investigating excited states of nuclei. An appealing feature of these
reactions is the clear understanding of the interaction mechanism. In
particular, electromagnetic dissociation (EMD) of halo nuclei has
revealed an anomalously large cross sections due to accumulation of
electric dipole (E1) strength at low energy. The first attempts to
interpret the phenomenon of low--energy excitations addressed a notion
of the so--called soft dipole
resonance~\cite{han87:4,kob89:232,suz91:528,esb92:542,ike92:538}.
However, it is still an open question whether this is a true resonant
state, which is being observed through the E1 excitation. For example,
intensive studies of the one--neutron halo nucleus \nuc{11}{Be} have
shown a very significant enhancement of the E1 strength just above the
\nuc{10}{Be}$\,+\,$n threshold where there are no resonances in
\nuc{11}{Be}. For \nuc{6}{He} several different three--body
approaches~\cite{suz91:528,dan93:302,cso94:49,fun94:575,dan98:632} agree
that there is no low--lying $1^-$ resonant state in the continuum. An
alternative way to explain the accumulation of E1 strength would be to
consider a direct breakup, where the final states are continuum states
without a resonant state. In the direct breakup mechanism, low--energy
dipole excitations occur as a natural consequence of the small binding
energy or correspondingly of the large distance between the charge and
the center of mass (CM) of a halo nucleus. In this way the E1 strength
function of the one--neutron halo nucleus \nuc{11}{Be} can be explained
using a simple two--body model with a Yukawa wave function (WF) for the
initial state and plane waves in the final
state~\cite{ber88:480,ots94:49}. However, for two--neutron halo nuclei,
like \nuc{6}{He}, \nuc{11}{Li}, and \nuc{14}{Be}, a three--body picture
is more appropriate but also more complicated. Microscopic
three--cluster calculations of the E1 strength function were performed
for, e.g., \nuc{6}{He}~\cite{dan98:632,cob98:58}. However, many
difficulties and questions caused by incomplete knowledge of the cluster
dynamics are still to be resolved. In the present paper we develop an
alternative approach. We construct phenomenological three--cluster bound
state WFs of two--neutron halo nuclei that behave correctly at large
intercluster distances, reproduce the nuclear sizes, and incorporate the
main features of the underlying three--body structure. No interaction
three--body WFs are used to describe the breakup final states. The model
allows an analytic calculation of the strength functions and can serve
as a helpful tool to predict the Coulomb disintegration spectra of a
variety of two--neutron halo nuclei. This can help to conduct new
experiments and eventually can provide a better understanding of the
microscopic structure of halo nuclei. An approach of this type has
previously been used by Pushkin \textit{et al.}~\cite{pus96:22} and the
present work can be viewed as a development of their model.

In the present paper the model is formulated. As a test case the E1
strength function of \nuc{6}{He} is studied and compared with the
experimental data of Aumann \textit{et al.}~\cite{aum99:59}. In a
subsequent paper the model will be applied to \nuc{11}{Li} and
\nuc{14}{Be}. In Sec. 2 our WFs are described, in Sec. 3 the strength
functions are obtained and discussed, and in Sec. 4 some conclusions are
presented. Details of the calculation are given in the Appendix.

For further reference we list here conventional formulae concerning the
EMD. We consider the case when an initial ground state is the only bound
state in a system (note that all known Borromean nuclei possess this
property) and all possible final states belong to the continuum.  In the
framework of first order perturbation theory the energy spectrum for E1
Coulomb excitation can be written as
\begin{equation}
  \frac{\d \sigma(\uE 1)}{\d E} = \frac{N_{\uE 1}(E^*)}{\hbar
  c}\frac{16 \pi^3}{9} \be.
\label{eq:xsec}
\end{equation}
Here $E^*$ is the excitation energy, $E^*=E_{0}+E$, where $E_{0}$ is the
binding energy, and $E$ is thus the continuum energy, $\d B(\uE1)/\d
E$ is the dipole strength function, and $N_{\uE 1}
(E^*)$~\cite{win79:319,ber85:442} is the spectrum of virtual photons
(where the actual number $\d n$ of virtual photons equals to
$N_{\uE 1} (E^*)(E^*)^{-1}\d E$). Since the spectrum of virtual
photons peaks at low energies, the Coulomb excitation to low--lying
states is favored as far as there exists a low--energy contribution to
the dipole strength.

The E1 strength function can be written as
\begin{equation}
\be=\frac{1}{2J_i + 1}
  \sum_{M_i}\sum_{\mu=-1,0,1}\int \d \tau_f\left| \langle f|
 \mathcal{M}(\uE 1,\mu) | i; J_i M_i \rangle \right|^2
  \delta\left(E_f-E\right).
\label{eq:strengthfuncdef}
\end{equation}
Here $\d \tau_f$ is the phase space element for final states,
$\vec{\mathcal{M}}(\uE 1)$ is the dipole operator, and $|i\rangle$,
$|f\rangle$ are the initial state and the final states in the CM
subsystem which are normalized as follows
\begin{equation}
  \langle i|i\rangle=1,\qquad \langle f|f'\rangle =\delta(\tau_f-\tau_f').
\label{eq:orth}
\end{equation}
In the case when discrete quantum numbers enter the labelling of
continuum final states the $\delta$--function notation adopted above
implies inclusion of $\delta$--symbols. Similarly, the notation $\int \d
\tau_f$ in Eq.~\eqref{eq:strengthfuncdef} may imply the inclusion of
summation over discrete quantum numbers.
  
Since halo nuclei exhibit a large degree of clusterization, low--energy
excitations will mainly affect relative motion between the $N$
clusters. The corresponding cluster E1 operator is
\begin{equation}
  \vec{\mathcal{M}}(\uE 1) = \sqrt{\frac{3}{4\pi}}
  \sum_{i=1}^N e Z_i (\vec{r}_i - \vec{R}_\ucm),
\label{eq:reldipoleop}
\end{equation}
where $\vec{r_i}$ are the cluster positions, and $\vec{R}_\ucm$ is the
position of the CM of the system. In the case of two--neutron halo
nuclei only the core will contribute in Eq.~\eqref{eq:reldipoleop}.

By summing the strength over all final states one obtains sum rules for
the process. In particular, the non--energy--weighted cluster sum rule
reads in our case as
\begin{equation}
  \int_0^\infty \frac{\d B (\uE 1)}{\d E} \d E =
  \frac{3}{4\pi} Z_\uc^2 e^2 \langle r_\uc^2 \rangle,
\label{eq:newcsr}
\end{equation}
where $Z_\uc$ is the charge of the core, $r_\uc$ is the distance between
the core and the CM of the whole system, and the average value is
calculated over the ground state WF.
\section{Wave functions and transition matrix elements}
In the present analysis we adopt the three--body model description of
two--neutron halo nuclei. The cluster part of the bound state WF, in the
CM subsystem, is written as an expansion over hyperspherical harmonics
(HH) (see e.g.~\cite{zhu93:231})
\begin{equation}
\Psi\left( \vec{x},\vec{y} \right) = \rho^{-5/2}
\sum_{KLSl_xl_y} \chi_{KLS}^{l_xl_y} \left( \rho \right)
\left[ \Gamma_{KL}^{l_xl_y} \left( \Omega_5 \right) \otimes \theta_S
\right]_{JM}.
\label{eq:hhwf}
\end{equation}
Here $\{\vec{x},\vec{y}\}$ is the set of Jacobi coordinates
\eqref{eq:normjac}, and $\{\rho,\Omega_5\}$ are the corresponding
hyperspherical coordinates. The quantity
$\rho=\left(x^2+y^2\right)^{1/2}$ is the hyperradius, and $\{\Omega_5\}$
denotes collectively five angles parametrizing a hypersphere with
$\rho=\mathrm{const}$. We use below that
\begin{equation}
\d\vec{x} \d \vec{y} = \rho^5 \d \rho \d \Omega_5.
\label{eq:vol}
\end{equation}
The HH, $\Gamma_{KLM_L}^{l_xl_y}\left( \Omega_5\right)$, form an
orthonormalized complete set. Harmonics with the orbital quantum numbers
$L,M_L$ are coupled with spin functions of two neutrons $\theta_{SM_S}$
to the total momentum $J,M$. The other quantum numbers labelling the HH
are the Jacobi orbital momenta $l_x$, $l_y$, and the hypermomentum $K$.
Since the WF~\eqref{eq:hhwf} should be antisymmetric with respect to the
valence neutrons it includes only terms with even $(l_x+S)$. More
details can be found in the Appendix.

For Borromean systems, having no bound subsystems, which include two
neutrons as constituents, the hyperradial functions entering the
expansion~\eqref{eq:hhwf} behave asymptotically as (see,
e.g.,~\cite{mer74:19})
\begin{equation}
\chi_\lambda (\rho) \rightarrow C_\lambda\exp(-\kappa_0 \rho), \qquad
\mathrm{as} \quad \rho \rightarrow \infty, 
\label{eq:asymp}
\end{equation}
where $\kappa_0$ is connected to the binding energy via $E_0 = (\hbar
\kappa_0)^2 /( 2 m)$, and $m$ is the nucleon mass. Thus, for a nucleus
with a small binding energy the WF has a long tail which is of
importance for peripheral reactions such as EMD. One could choose a
phenomenological bound state WF using the normalized hyperradial
function
\begin{equation}
\chi^{(1)} (\rho) \equiv \sqrt{2\kappa_0} \exp(-\kappa_0 \rho),
\label{eq:1pwf}
\end{equation}
together with the HH from expansion~\eqref{eq:hhwf} which has a
predominant weight. The single free parameter, $\kappa_0$, is fixed from
the binding energy. Such a model WF, incorporating the $K=0$ HH, has
been used in Ref.~\cite{pus96:22} for the \nuc{11}{Li} case and led to
an analytic expression for the E1 strength function reproducing well the
shape of existing experimental data. The total WF behaves as
$\rho^{-5/2}$ for small $\rho$ but it is still normalizable since the
singularity cancels with the $\rho^5$ factor in the volume element,
Eq.~\eqref{eq:vol}. However, using this model the WF is overestimated at
small $\rho$ and thus the $\sqrt{2\kappa_0}$ value is smaller than the
true asymptotic constant. This leads to an underestimation of
$\langle\rho^2\rangle$ and consequently to an underestimation of the
size of the system since, in accordance with Eq.~\eqref{eq:rms},
\begin{displaymath}
\langle R^2_\urms\rangle=A^{-1}\left[\langle\rho^2\rangle+(A-2)\langle 
R^2_\urms(\mathrm{core})\rangle\right],
\end{displaymath}
where $A$ is the mass number, and the last term represents the intrinsic
size of the core.

In our model, to cure this feature, we will be add an extra hyperradial
term of the same exponential form to reproduce simultaneously the true
asymptotic behavior of the ground state (connected to the binding
energy) and its size. The corresponding normalized function is
\begin{equation}
\begin{split}
\chi^{(2)} (\rho) & \equiv c\left[ \exp(-\kappa_0 \rho) -
\exp(-\kappa_1 \rho) \right], \\ 
& \mathrm{where} \quad c= \sqrt{\frac{2 \kappa_0 \kappa_1
(\kappa_0 + \kappa_1)} {(\kappa_0 - \kappa_1)^2}}.
\label{eq:2pwf}
\end{split}
\end{equation}
The parameters $\kappa_0$ and $\kappa_1$ are fixed using experimental
values of binding energy and rms radius. With the condition that
$\kappa_1 > \kappa_0$ we ensure that the second term decays faster than
the first, and thus the correct asymptotics is preserved. Furthermore,
the divergence of the total WF at small $\rho$ has been reduced to
$\rho^{-3/2}$. The particular form of the function~\eqref{eq:2pwf} has
been chosen to be able to perform the calculations analytically.

We shall retain only one or a few terms in the HH
expansion~\eqref{eq:hhwf}. This approximation is motivated by the rapid
increase of the multidimensional centrifugal barrier as the hypermoment
$K$ increases, see e.g.~\cite{zhu93:231}. The predominant terms in the
expansion usually correspond to the lowest possible value of $K$ that is
not suppressed by the Pauli principle.

In the \nuc{6}{He} case the above approximations lead us to the
following normalized, model WF for the initial bound state ($J^\pi=0^+$)
\begin{equation}
\begin{split}
\Psi(\vec{x},\vec{y}) = \frac{\chi^{(N)} (\rho)}{\rho^{5/2}}
\left\{ \left[ \sqrt{w_{00}} \Gamma_{000}^{00} \left( \Omega_5 \right) +
\sqrt{w_{20}} \Gamma_{200}^{00} \left( \Omega_5 \right) \right]
\theta_{00} \right.\\
\left. + \sqrt{w_{21}} \left[ \Gamma_{21}^{11} \left( \Omega_5 \right)
\otimes \theta_1 \right]_{J=0} \right\}, \qquad \mathrm{where~} w_{00} +
w_{20} + w_{21} =1.
\label{eq:modelwf}
\end{split}
\end{equation}
Here $\chi^{(N)}$ is given by Eq.~\eqref{eq:2pwf} (or alternatively by
Eq.~\eqref{eq:1pwf}). For the HH retained in Eq.~\eqref{eq:modelwf}
realistic hyperradial functions, in the corresponding expansion, behave
rather similarly at large $\rho$ of interest, so that the model with
hyperradial functions of the same form is permissible. The $\alpha$--N
interaction is repulsive in s states due to the Pauli principle, and it
is attractive in p states. Because of this the $K=0$ contribution to the
WF of \nuc{6}{He} is small, and the lowest $K$ value, which is not Pauli
suppressed, is $K=2$. From the $K=2$ contributions to the WF, given by
the second and third terms in Eq.~\eqref{eq:modelwf}, only the
$\Gamma_{200}^{00}$ term involves the s wave NN attraction and at the
same time it includes a very large component with the p wave $\alpha$--N
attraction. As a result, the term with $\Gamma_{200}^{00}$ should have a
predominant weight, and to a first approximation we may keep only this
single term. One can also retain all three terms entering
Eq.~\eqref{eq:modelwf} using stable estimates for the weights of the HH
components taken from theoretical predictions~\cite{zhu93:231}, see
Table~\ref{tab:wfs}. Similar phenomenological WFs may be constructed
for other two--neutron halo nuclei. Explicit expressions for the HH
entering the initial state \eqref{eq:modelwf} are given in
Eq.~\eqref{eq:ishh}.

In Table~\ref{tab:wfs} parameters of our \nuc{6}{He} bound state WFs
thus obtained are listed. The experimental values of the two--neutron
separation energy $S_{2n} = 0.97$~MeV and the rms radius $R_\urms =
2.50$~fm are adopted for $\Psi_2$ and $\Psi_3$. The $R_\urms=1.79$ fm
for the function $\Psi_1$ is too small as discussed above.

In Fig.~\ref{fig:microwf}(a) our model hyperradial
function~\eqref{eq:2pwf} is compared with the hyperradial functions
obtained from microscopic three--body calculations of
\nuc{6}{He}~\cite{zhu93:231}. (For other halo nuclei such functions are
not known with a sufficient confidence.) Fig.~\ref{fig:microwf}(b)
illustrates the ranges of $\rho$ values important for calculating the
strength function at energies not far away from its maximum. Typical
integrands entering the matrix elements (ME), Eq.~\eqref{eq:int}, are
plotted. They include our hyperradial bound state function and
hyperradial components of final state WFs. In our case the latter
components correspond to $K=1$ and $K=3$ (cf. below) from which the
$K=1$ case is shown. The plot for $K=3$ looks similar. One can see that
at energies not too far from the peak of the strength function, at $E
\sim 2$~MeV~\cite{aum99:59}, the model hyperradial function is on
average rather close to the realistic ones at $\rho$ values of
interest. (We note in this connection that in the peak region the
contribution of the $K=0$ component of the WF to the net result is much
higher than its relative weight.) At energies in the maximum region our
model ground state WF leads to a strength function which is very close
to that obtained with the realistic ground state WF. Away from the
maximum region one should expect more strength with our model than with
the realistic ground state WF provided that the final state WFs are the
same.

We calculate the strength function disregarding the final state
interaction (FSI). We shall see that a reasonable agreement with
experiment both in the energy dependence of the strength function and in
its magnitude emerges in this way.

To obtain a differential cross section of EMD one would have to use
final state WFs with given momenta, including angular information. When
the FSI is disregarded these WFs are three--body plane waves. To carry
out the calculations these plane waves could be expanded in products of
HH in coordinate and momentum spaces, see
e.g.~\cite{for00:673}. However, in our inclusive case, when we are only
interested in the energy dependence of the cross section, we do not need
directions of the momenta. Thus, instead of using plane waves, we will
use a set of no interaction final states that include just the
coordinate space HH. Indeed, the strength
function~\eqref{eq:strengthfuncdef} does not depend on the choice of
final state set (for the same Hamiltonian) provided that the proper
orthonormality conditions~\eqref{eq:orth} are fulfilled. The states we
choose are labelled by energy, hyperspherical quantum numbers
$K,L,l_x,l_y$, and quantum numbers $J,M,S$ of total momentum and of spin
of the valence neutrons.  The corresponding configuration space WFs
include Bessel functions, and they are of the form
\begin{equation}
\frac{J_{K+2}(\kappa\rho)}{(\kappa\rho)^2}
\left[ \Gamma_{KL}^{l_xl_y} \left( \Omega_5 \right) \otimes \theta_S
\right]_{JM}.
\label{eq:cwf}  
\end{equation}
Alike the plane waves these functions are solutions to the free--space,
six--dimensional Schr{\"o}dinger equation. The continuum energy is
related to $\kappa$ via $E_f =( \hbar \kappa)^2 /( 2 m)$, see
e.g.~\cite{for00:673}. The phase space element needed in
Eq.~\eqref{eq:strengthfuncdef} to integrate over the contributions of
states~\eqref{eq:cwf} is
\begin{equation}
\int\d\tau_f=\sum_{Kl_xl_yLSJM}\int \kappa^5\d\kappa=
4 \left( \frac{m}{\hbar^2} \right)^3
\sum_{Kl_xl_yLSJM}\int E_f^2 \d E_f 
\label{eq:el}
\end{equation}
where the sum is over final states. In accordance with
Eqs.~\eqref{eq:orth} and~\eqref{eq:el} the states~\eqref{eq:cwf} are
normalized to $\kappa^{-5}\delta(\kappa-\kappa')$ times the
$\delta$--symbol with respect to the discrete quantum numbers.

The transition ME between components of the initial and final states
with given $K$ values obey the selection rule $\Delta K = \pm 1$ (the
reason is that the transition dipole operator is a HH with $K=1$). HH
have parities $(-1)^K$ so that, e.g., in the usual case when the cluster
part of the ground state WF has positive parity, final states
\eqref{eq:cwf} with odd $K$ contribute to the result. Taking
Eqs.~\eqref{eq:vol},~\eqref{eq:2pwf} and~\eqref{eq:modelwf} into account
together with the proportionality of the transition operator to $\rho$,
the ME of $\vec{\mathcal{M}}(\uE 1)$ operator~\eqref{eq:reldipoleop}
will include hyperradial integrals of the form
\begin{equation}
(\kappa/\kappa_i)^{-2}\int_0^\infty J_{K+2}((\kappa/\kappa_i)z)z^{3/2}
\e^{-z}\d z,
\label{eq:int}
\end{equation}
with $i=0$ and 1. Note that the whole energy dependence of the total ME
is included in these integrals.
\section{E1 strength functions for two--neutron halo nuclei} 
\begin{figure}[tbh]
  \begin{center}
  \begin{minipage}{0.95\textwidth}
    \begin{minipage}[t]{0.47\textwidth}
      \centering
      \includegraphics[width=\textwidth]{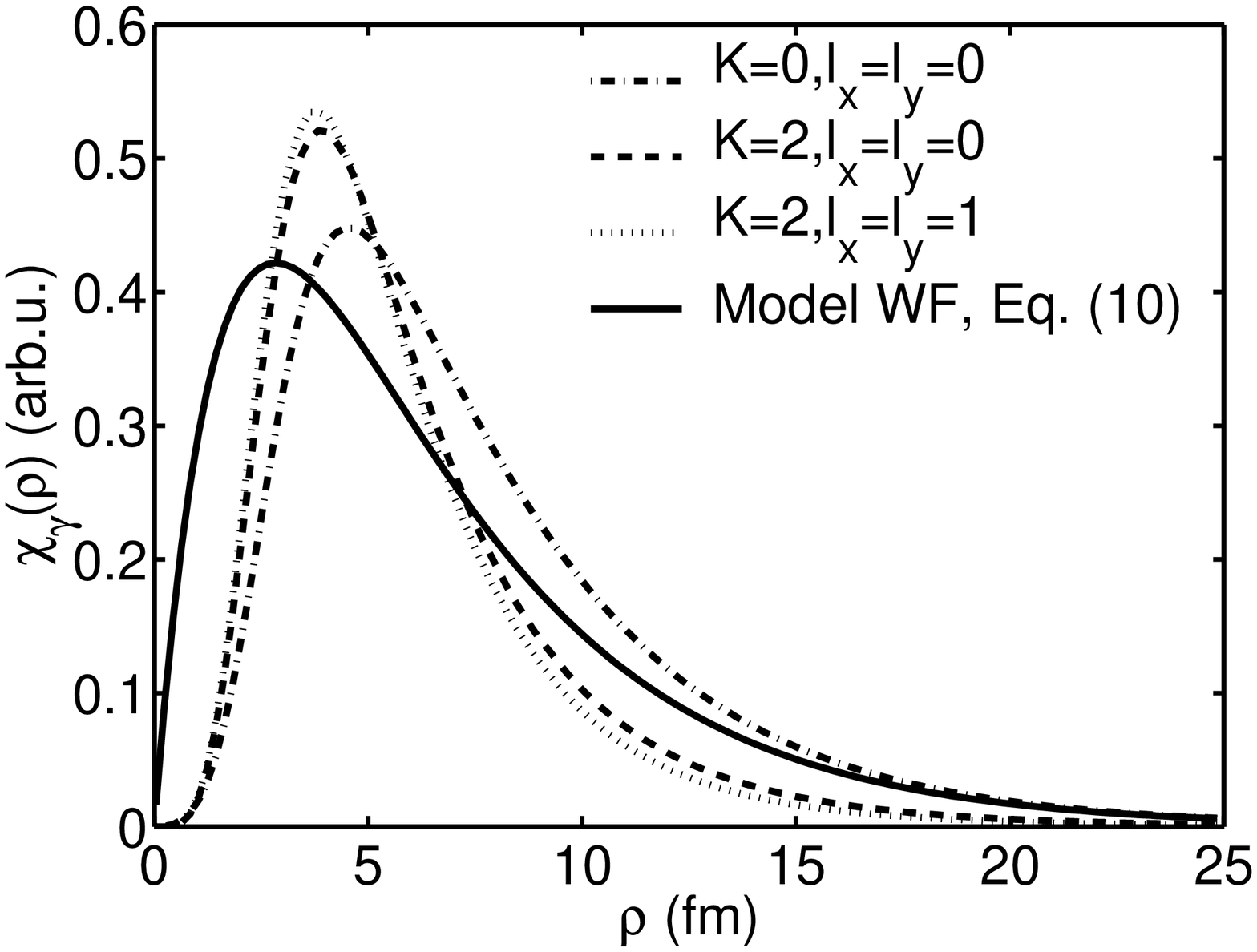}
    \end{minipage}
  \hfill
    \begin{minipage}[t]{0.47\textwidth}
      \centering
      \includegraphics[width=\textwidth]{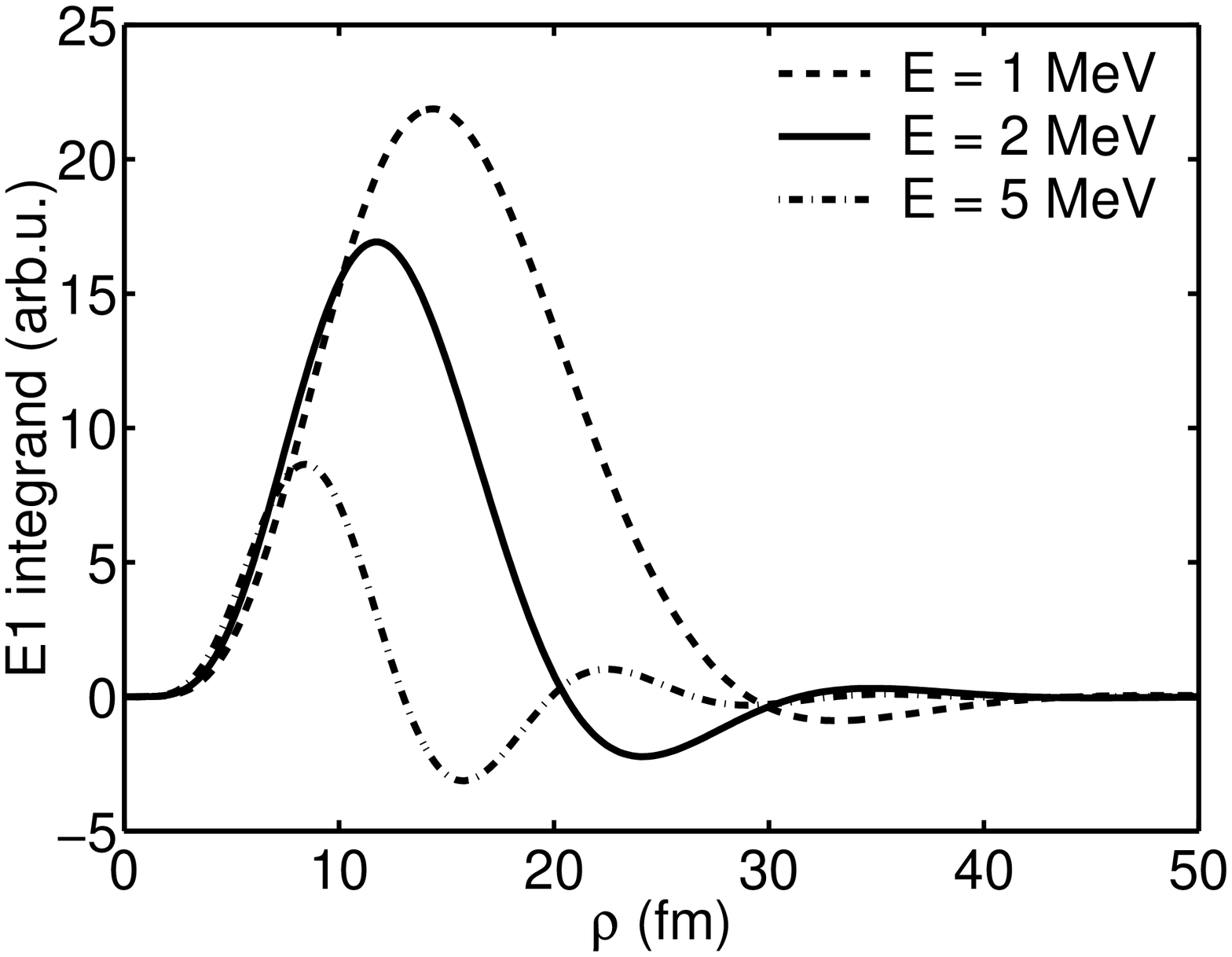}
    \end{minipage}
    \begin{minipage}[b]{0.47\textwidth}
      \centering
      (a)
    \end{minipage}
  \hfill
    \begin{minipage}[b]{0.47\textwidth}
      \centering
      (b)
    \end{minipage}
  \end{minipage}
  \caption{(a) The model hyperradial function $\chi^{(2)}$,
  Eq.~\eqref{eq:2pwf} (solid line), compared with the three predominant
  hyperradial functions in expansion~\eqref{eq:hhwf} obtained
  from a microscopic three--body calculation of
  \nuc{6}{He}~\cite{zhu93:231}. All functions have been normalized to
  unity. (b) Typical integrands entering the E1
  transition ME~\eqref{eq:int}. This plot illustrates the important
  range of $\rho$ values.%
    \label{fig:microwf}}
  \end{center}
\end{figure}

Before presenting more detailed results we shall obtain an approximate
general property of the strength functions. Let us compare to each other
the strength functions for a pair of two--neutron halo nuclei. We assume
first that the initial ground states of both nuclei are dominated by the
HH with $K=0$. Then the largest contributions to the strength functions
come from the final states with $K=1$. At moderate energies predominant
contributions to our ME come from high $\rho$ values. For time being let
us suggest that the $\rho$ dependence of the ground state WFs at such
$\rho$ values is described by Eqs.~\eqref{eq:hhwf},~\eqref{eq:asymp}. As
a result the energy dependencies of both strength functions will mainly
be determined by the quantities \eqref{eq:int} with $i=0$ and
$K=1$. Thus these energy dependencies differ only in the parameter
$\kappa_0$. We come to the conclusion that if the strength functions of
different two--neutron halo nuclei are plotted on the scales $E/E_0$
their forms should be similar to each other. This feature is
characteristic of the no--FSI approximation but, in view of the results
below, one can hope that it remains approximately valid also beyond this
approximation. Another limitation is that in reality the high $\rho$
behaviour~\eqref{eq:asymp} may not (in some cases) be achieved at $\rho$
values of interest, and power corrections in the corresponding expansion
\begin{displaymath}
C_\lambda\exp(-\kappa_0\rho)(1+a\rho^{-1}+b\rho^{-2}\ldots)
\end{displaymath}
can still be important. Some of these corrections violate the scaling
property. However, within the ranges of $\rho$ values effectively
contributing to~\eqref{eq:int}, see Fig.~\ref{fig:microwf}(b), the
change of the power terms is relatively small as compared to that of the
exponential and to a certain approximation the scaling property remains
valid.

This property should hold true also when the ground states are dominated
by a HH with $K \ne 0$ which is the same for both nuclei, say
$\Gamma_{200}^{00}$ with $K=2$. In such a case the same combinations of
hyperradial integrals~\eqref{eq:int} will define the strengths. The
scaling property also fulfills in all cases when the contribution of the
hyperradial integrals~\eqref{eq:int} with $K=1$ dominates the strengths.
 
\begin{figure}[tbh]
  \begin{center}
  \includegraphics[width=0.6\textwidth]{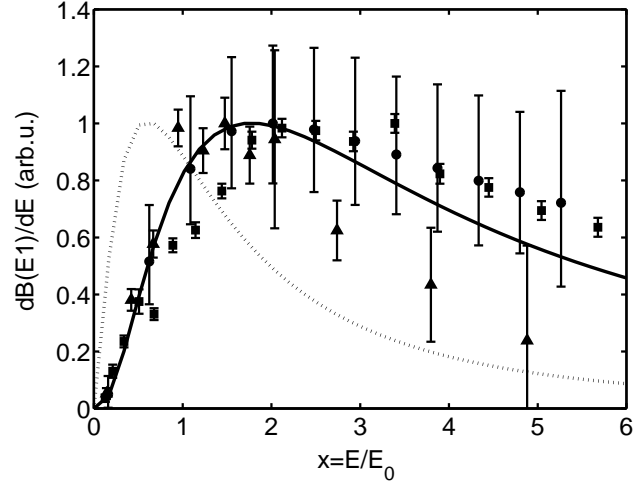}
  \caption{Comparison of scaled, experimental strength functions for
  \nuc{6}{He} and \nuc{11}{Li}. The experimental data are: (circles)
  \nuc{6}{He} -- Aumann~\textit{et~al.}~\cite{aum99:59}, (triangles)
  \nuc{11}{Li} -- Sackett~\textit{et~al.}~\cite{sac93:48}, (squares)
  \nuc{11}{Li} -- Shimoura~\textit{et~al.}~\cite{shi95:348}. The
  dotted curve is the two--body strength function from
  Ref~\cite{ber88:480,ots94:49}. The solid line is the three--body
  strength function from Eq.~\eqref{eq:str}.%
  \label{fig:scaledexp}} \end{center}
\end{figure}

In Fig.~\ref{fig:scaledexp} several experimentally measured strength
functions for \nuc{6}{He} and \nuc{11}{Li} are compared in the described
way. The spectra are plotted as functions of the parameter $x = E /
E_0$. The similarity of these scaled strength functions is evident. Such
plots may be useful for quick predictions of shapes of E1 strength
functions for two--neutron halo nuclei or for checking hypotheses on
their three--body structure. Shown in this figure is also the analytical
strength function from a two--body calculation where a Yukawa WF has
been used together with plane waves in the final
state~\cite{ber88:480,ots94:49}. The two--body model gives a peak at too
low energy thus clearly indicating the importance of using a three--body
approach.

Now let us obtain the E1 strength function of
Eq.~\eqref{eq:strengthfuncdef} in our model. The final states
\eqref{eq:cwf} which can be reached from the ground state
\eqref{eq:modelwf} via the dipole excitation are those with $K=1$ and
$K=3$. Other selection rules are $\Delta l_x = 0$, $\Delta l_y =\pm 1$,
and $\Delta L=0,\pm 1$. The HH entering the final states~\eqref{eq:cwf}
that contribute to the result for \nuc{6}{He} are listed in
Eq.~\eqref{eq:fshh}. The required ME are sums of products of
hyperangular ME and hyperradial integrals. The hyperangular ME entering
the calculation are listed in Eq.~\eqref{eq:hme}. The hyperradial
integrals are of the form~\eqref{eq:int}, they can be calculated
analytically and expressed (see e.g.~\cite{gra80}) in terms of the
hypergeometrical functions. When the model~\eqref{eq:2pwf},
\eqref{eq:modelwf} is adopted for the bound state, one obtains the
following final expression for the E1 strength function:
\begin{equation}
\be=c^2 D E^3\sum_{i,j=0}^1
\frac{(-1)^{i+j}
\left[\alpha_1 F_1(y_i)F_1(y_j)+
\alpha_2F_2(y_i)F_2(y_j)\right]}{\left[ (E_i+E) (E_j+E)
    \right]^{11/4}}.
\label{eq:str}
\end{equation}
Here $E$ is the continuum energy, $E_{0,1}=(\hbar\kappa_{0,1})^2/(2m)$,
where $\kappa_0$, $\kappa_1$ and $c$ are defined in
Eq.~\eqref{eq:2pwf}. The constant $D$ is
\begin{displaymath}
D=\frac{3}{2} \left( \frac{\hbar^2}{2m}
    \right)^{3/2} \frac{(Z_\uc e)^2}{(A- 2)A},
\end{displaymath}
and the constants $\alpha_1$ and $\alpha_2$ are   
\begin{displaymath}
  \alpha_1 = \frac{1}{4} \left( \frac{315}{2^{10}} \right)^2 \left( 1 +
  3w_{00} + 4\sqrt{w_{00}w_{20}} \right),\qquad
  \alpha_2 = \left( \frac{9009 \sqrt{3}}{2^{17}} \right)^2 \left( 
  w_{20} + w_{21} \right).
\end{displaymath}
The coefficients $w_{00}$, $w_{20}$, and $w_{21}$ entering here are 
the weights of various HH, see Eq.~\eqref{eq:modelwf}. The quantities
$y_{0,1}$ are defined as $y_i=E/(E+E_i)$, and, finally,
\begin{equation}
 F_1(y)=F\left( \frac{11}{4}, \frac{3}{4}, 4,y\right),\qquad
F_2(y)=yF \left( \frac{15}{4}, \frac{7}{4}, 6,y\right),
\label{eq:gf}
\end{equation}
where $F(\alpha, \beta, \gamma, z)$ is the standard hypergeometrical
function~\cite{gra80}. The functions~\eqref{eq:gf} represent the
contributions from final states with $K=1$ and $K=3$, respectively. When
the hyperradial function~\eqref{eq:1pwf} is used the result is obviously
obtained from Eq.~\eqref{eq:str} by retaining only the term with $i=j=0$
and replacing $c^2$ with $2\kappa_0$. At $E\rightarrow 0$ the strength
function~\eqref{eq:str} exhibits the typical $E^3$ three--body behavior
at threshold. (In reality ``dineutron'' correlations may influence the
spectrum in the threshold region which is beyond our present
consideration.)
 
\begin{figure}[tbh]
  \begin{center}
  \begin{minipage}{0.95\textwidth}
    \begin{minipage}[t]{0.47\textwidth}
      \centering
      \includegraphics[width=\textwidth]{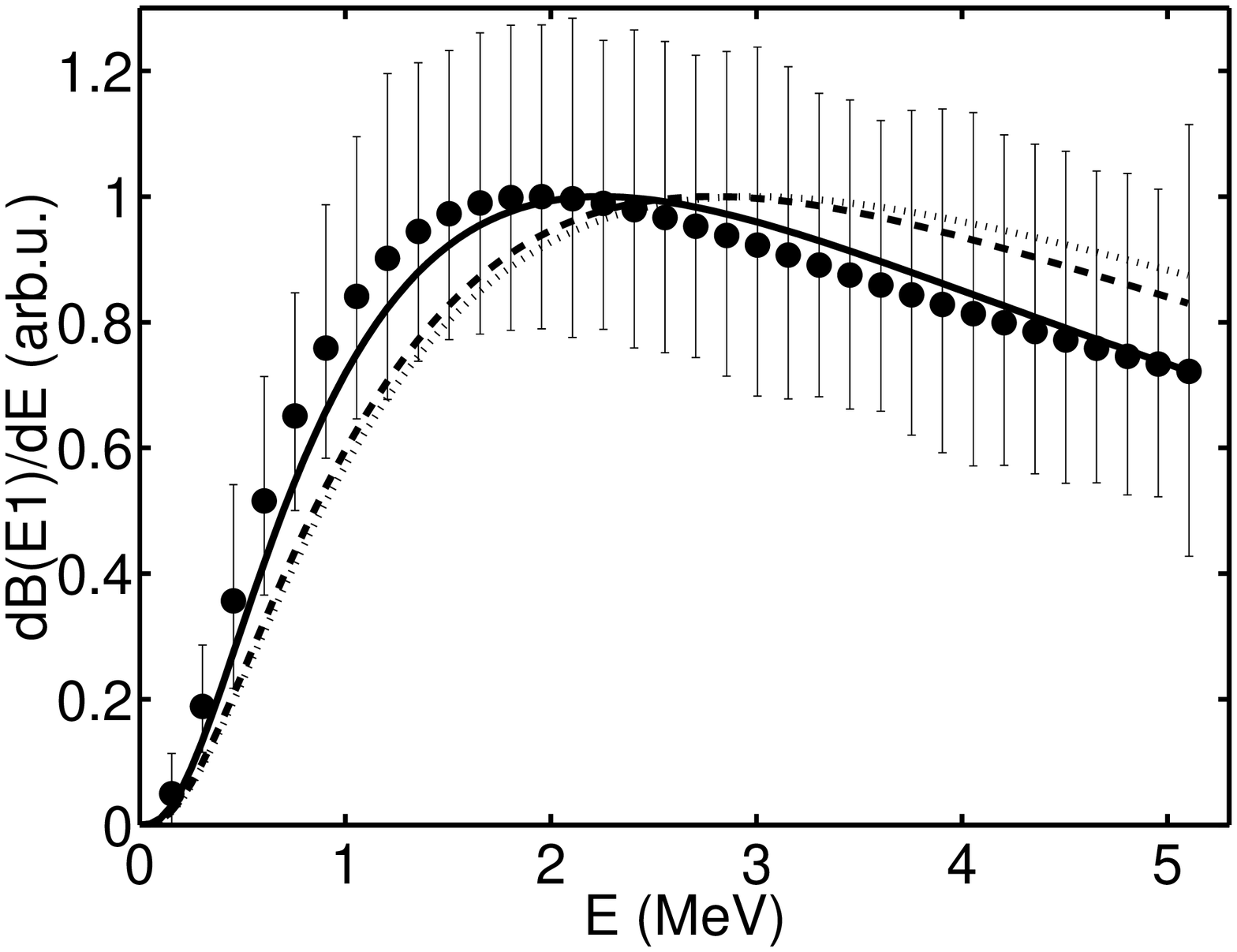}
    \end{minipage}
  \hfill
    \begin{minipage}[t]{0.47\textwidth}
      \centering
      \includegraphics[width=\textwidth]{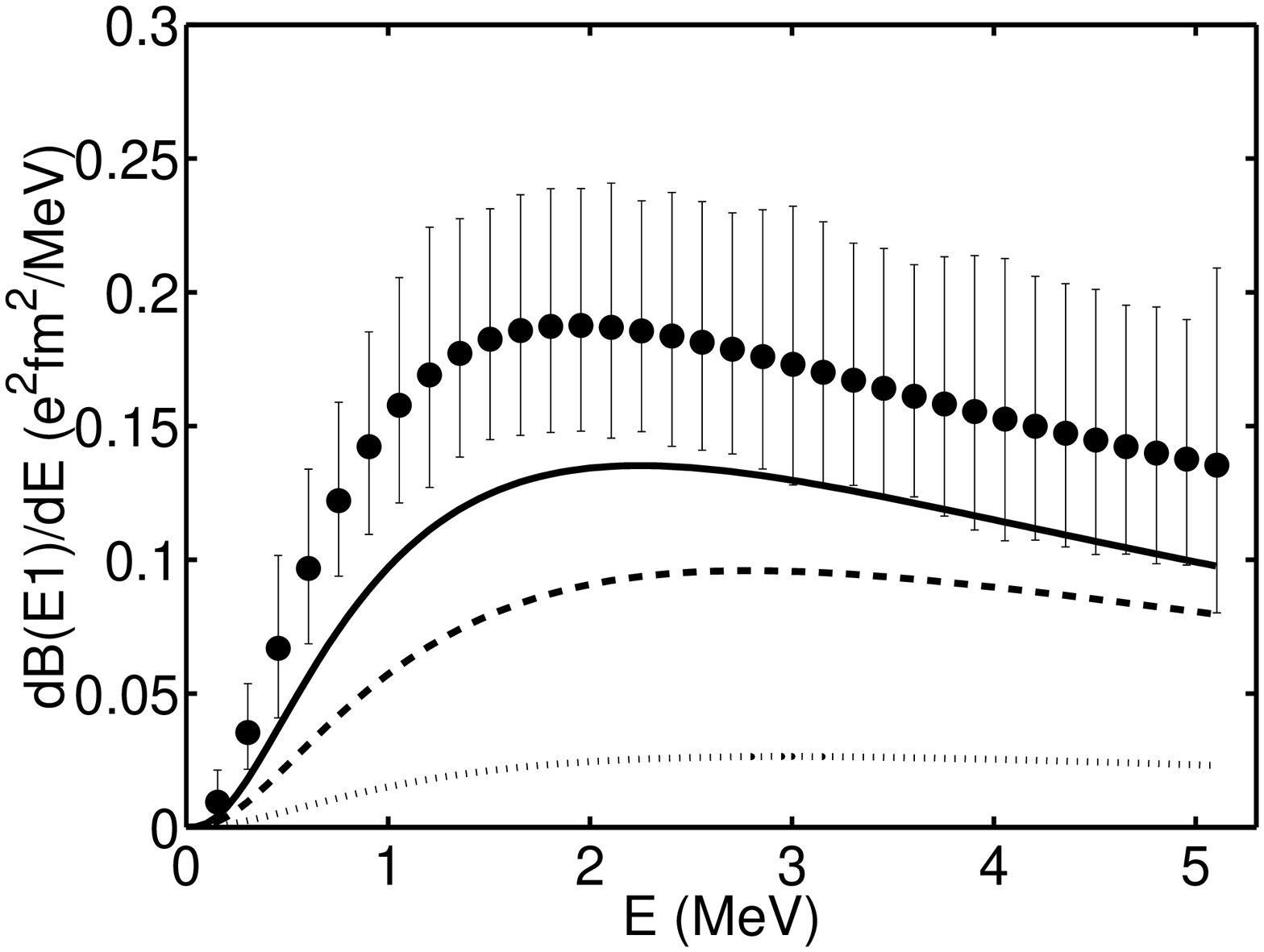}
    \end{minipage}
    \begin{minipage}[b]{0.47\textwidth}
      \centering
      (a)
    \end{minipage}
  \hfill
    \begin{minipage}[b]{0.47\textwidth}
      \centering
      (b)
    \end{minipage}
  \end{minipage}
  \caption{A comparison of our strength function with experimental data
  for \nuc{6}{He}, Aumann~\textit{et~al.}~\cite{aum99:59}, both in
  arbitrary units \textnormal{(a)} and in absolute scale
  \textnormal{(b)}. The analytical strength functions are obtained with
  WFs $\Psi_{1}$ (dotted), $\Psi_{2}$ (dashed) and $\Psi_{3}$
  (solid), see Table~\ref{tab:wfs}.%
    \label{fig:he6strength}}
  \end{center}
\end{figure}

In Fig.~\ref{fig:he6strength} experimental data for the \nuc{6}{He}
strength function from Aumann \textit{et al.}~\cite{aum99:59} (Pb
target, $E_\mathrm{lab}=240$ MeV/nucleon) are compared to our model
strength functions calculated with Eq.~\eqref{eq:str}. The comparison is
done both for the energy behavior of the spectra,
Fig.~\ref{fig:he6strength}(a), and in absolute scale,
Fig.~\ref{fig:he6strength}(b). From the comparison in
Fig.~\ref{fig:he6strength}(a) one sees that the shape of the spectrum is
well reproduced by our model, especially in the case of the most
complete version $\Psi_3$, see Table~\ref{tab:wfs}. A shift of the
maximum to lower energies in the latter case is related to the inclusion
of the contribution from the ground state HH with $K=0$.  Maxima of the
contributions from the HH with $K=2$ are shifted to higher energies
since these HH lead in part to excitation of $K=3$ final states.

An interesting feature of the calculated strength functions seen in
Fig.~\ref{fig:he6strength}(a) is that when one passes from the
hyperradial function~\eqref{eq:1pwf} to the function~\eqref{eq:2pwf}
retaining the same hyperangular dependence (e.g. going from $\Psi_1$ to
$\Psi_2$) the shape of the spectrum changes very little. This is
explained by the above mentioned fact that only large $\rho$ values
contribute sizably to the result. For such $\rho$ values only the first,
longer range exponential in the hyperradial function~\eqref{eq:2pwf}
survives leading to just the same energy dependence as the
model~\eqref{eq:1pwf} (but to very different magnitudes of the cross
section). Indeed, if one plots a figure similar to
Fig.~\ref{fig:microwf}(b) but retaining only the contribution from the
first exponential to the ME one obtains the integrands extremely close
to those shown in Fig.~\ref{fig:microwf}(b). The same holds true for the
$K=3$ final state. This means that the $i=j=0$ term dominates
Eq.~\eqref{eq:str} at not too high energies. The other terms are
suppressed by the factor $[(E_0+E)/(E_1+E)]^{11/4}$. At higher energies
the model $\Psi_2$ leads to a strength function which decreases faster
than that for the model $\Psi_1$. The internal part of the hyperradial
function contributes more to the results here, and in the case of the
two--parameter model this internal part is smaller than for the
one--parameter model.
 
Comparing the magnitudes of the calculated strength functions with
experiment in Fig.~\ref{fig:he6strength}(b) one sees that the ground
state model $\Psi_1$ (see Table~\ref{tab:wfs}) leads to a very low
strength function. The results are considerably improved when passing to
the model $\Psi_2$ while the model $\Psi_3$ leads to a further
improvement and compares reasonably well with experiment. These results
may be commented as follows. Underestimating the $\rho_\urms$ value the
one--parameter model $\Psi_1$ underestimates simultaneously all sizes in
the system including the distance $r_\uc$ between the core and the CM,
see Table~\ref{tab:wfs}. Furthermore, one needs to take into account
that the E1 sum rule~\eqref{eq:newcsr} strictly preserves its value if
one replaces true final states with a complete set of no interaction
final states. So, in our case the total strength is determined by the
$r_\uc$ value, and thus it is natural that the above model leads to a
too low total strength. While the two--parameter WF, $\Psi_2$,
reproduces correctly the rms matter radius it still gives an $r_\uc$
value which is somewhat low. For this WF, which includes only one HH
with $l_x=l_y$, we have the simple relation $\langle x^2\rangle =
\langle y^2 \rangle = \langle \rho^2 \rangle /2$. However, this relation
breaks down, due to off--diagonal terms in the expectation values, when
more HH are included, as in $\Psi_3$, and the geometric properties of
the state change slightly leading to a higher $r_\uc$ value.

In accordance with the above discussion, at not too high energy the main
effect of passing from the one--parameter model~\eqref{eq:1pwf} to the
two--parameter model~\eqref{eq:2pwf} consists merely in the
multiplication of the strength function by $c^2/(2\kappa_0)\simeq 4$
which can be seen from comparing the strength functions for models
$\Psi_1$ and $\Psi_2$ in Fig.~\ref{fig:he6strength}(b). This is in
principle similar to the case of photodisintegration of
deutrons~\cite{lev60}. The cross section, calculated with the
zero--range Yukawa ground state wave function
$(2\kappa_0)^{1/2}\exp(-\kappa r)/r$ and the no interaction final state,
is lower than the experimental one. The main effect of the finite range
correction consists in an increase of the asymptotic constant in the
ground state WF leading to a reasonably good comparison with experiment
at moderate energy.

In Table~\ref{tab:he6strength} the integrated strengths obtained are
listed along with the experimental strengths and those obtained from a
microscopic three--cluster calculation. The strengths are integrated up
to $E=4$ MeV, $9$ MeV, and infinity. Note that for the first of these
energies the ratio of the result obtained with $\Psi_2$ to that obtained
with $\Psi_1$ is about $c^2/(2\kappa_0)$ as it should be. Our total
strength proved to be very close to that obtained using the microscopic
bound state WF. These results on the total strength provide us also with
a check of the strength function calculation. Integrating the strength
function up to infinite energy one should reproduce the value of the sum
rule~\eqref{eq:newcsr}.

\begin{figure}[tbh]
  \begin{center}
  \includegraphics[width=0.6\textwidth]{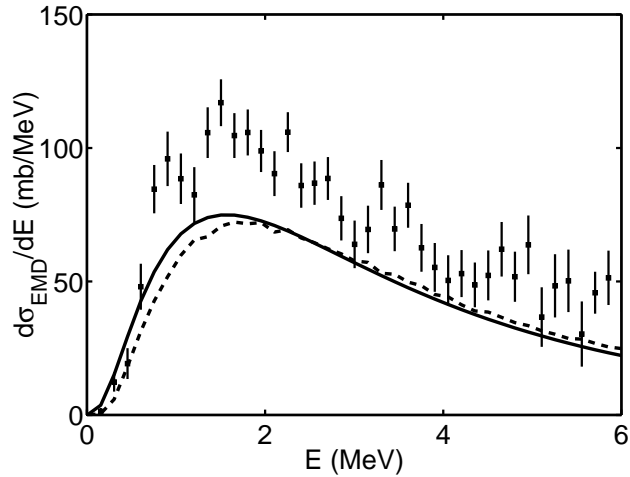}
  \caption{The \nuc{6}{He} EMD excitation energy spectrum  obtained at
  240~MeV/A with a lead target. The data points are extracted from
  Ref.~\cite{aum99:59} while the solid line is our result obtained by
  multiplying the strength function~\eqref{eq:str} calculated for the
  $\Psi_{3}$ model with the spectrum of virtual photons. The dashed line
  is the same curve corrected for the LAND detector response and
  efficiency.%
  \label{fig:he6xsec}} \end{center}
\end{figure}

Finally, we calculate the EMD excitation energy spectrum using 
Eq.~\eqref{eq:xsec}. The virtual photon spectrum was obtained for a
\nuc{6}{He} beam with energy 240~MeV/A striking a Pb target. In
Fig.~\ref{fig:he6xsec} the resulting distribution is compared with the
excitation energy spectrum measured by Aumann \textit{et
al.}~\cite{aum99:59}. In Table~\ref{tab:he6xsec} the total \nuc{6}{He}
cross sections obtained with our model are compared with experimental
estimates for two different beam energies, 37~MeV/A and 240~MeV/A,
measured at NSCL~\cite{war00:62} and GSI~\cite{aum99:59} respectively.
The theoretical cross section has been calculated by integrating the
spectrum up to the energy $E=10$~MeV, corresponding approximately to
the maximum energy measured in the experiments. The observed
difference should be a direct measure of the importance of FSI since the
virtual photon spectrum peaks at low energies and this is the region
were FSI is supposed to play a role. We find that our model gives $\sim
35 \%$ lower cross section than experiment for both energies.
\section{Conclusion}
A three--body model describing EMD of two--neutron halo nuclei, which
allows studying the E1 strength functions without precise knowledge
about initial and final states, has been developed.  The model leads to
an analytical expression for the strength functions.  Our ground state
wave functions reproduce both the true three--body asymptotics
(determined by the binding energy) and the correct size of the system.
A complete set of three--particle no interaction final states is
used. We found that the large distance asymptotics of the ground state
determines the shape of the strength function, while the size of the
ground state governs the asymptotic constant and thus the magnitude of
the strength function. We pointed out that in some cases the shapes of
the E1 strength functions for different two--neutron halo nuclei are
approximately related to each other.

We have used \nuc{6}{He} as a test case and made an extensive comparison
with experimental data~\cite{aum99:59,war00:62}.  We found a good
agreement concerning the shape and peak position of the E1 strength
function and a reasonable agreement concerning its magnitude.

A remarkable feature of halo nuclei is that the E1 strength is
concentrated at low energies even without any low--lying resonant
state. This peculiarity is due to the low binding energy and large size
of the initial state which is demonstrated by the present calculation
with no final state interaction included in the model.

In a forthcoming paper, see also~\cite{for00:lic}, we shall apply the
present approach to investigate the E1 strength functions of Borromean
\nuc{11}{Li} and \nuc{14}{Be} nuclei. Common for these nuclei are large
uncertainties concerning their microscopic structure. In the
\nuc{11}{Li} case the E1 strength function is not so well known
experimentally as in the \nuc{6}{He} case, and for \nuc{14}{Be} the EMD
process has been measured only very recently~\cite{lab01:86}.

As to the FSI effects, we noted that the total E1 strength will not
change when replacing a complete set of no interaction final states with
such a set of true final states. Therefore the only effect of including
FSI will be a redistrubution of the strength leading to a somewhat
higher strength at low energy.  Further studies on this point would be
of interest.
\clearpage
{\appendix
\section{Appendix}
%
\subsection{Coordinate sets and hyperspherical harmonics}
%
We need to perform the calculation in the CM subsystem.  
The adopted Jacobi coordinates are 
\begin{equation}
\begin{split}
    \vec{x} &= \frac{1}{\sqrt{2}} (\vec{r}_1-\vec{r}_2), \\
    \vec{y} &= \sqrt{ \frac{2(A-2)}{A} } \left(
      \frac{\vec{r}_1+\vec{r}_2}{2} - \vec{r}_\uc \right),   
  \label{eq:normjac}
\end{split}
\end{equation}
where $\vec{r}_1$, $\vec{r}_2$, and $\vec{r}_\uc$ are positions of the
valence nucleons and the core. We use the related hyperspherical
coordinates $\{\rho,\Omega_5\}$.  The hyperradius $\rho$ determines the
size of a three--body state:
\begin{equation}
\rho^2=x^2+y^2=(\vec{r}_1-\vec{R}_\ucm)^2+(\vec{r}_2-\vec{R}_\ucm)^2
+(A-2)(\vec{r}_\uc-\vec{R}_\ucm)^2.
\label{eq:rms}
\end{equation}
The five angles $\{\Omega_5\}$ include usual angles $(\theta_x,\phi_x)$,
$(\theta_y,\phi_y)$ parametrizing the unit vectors $\hat{\bf x}$,
$\hat{\bf y}$ and the hyperangle $\theta$ defined by the equalities
\begin{equation}
  x = \rho\sin\theta,\qquad
  y = \rho\cos\theta,
 \label{eq:theta}
\end{equation}
where $0 \le \theta \le \pi/2$. The volume element $d\Omega_5$ entering
\eqref{eq:vol} is
\begin{equation}
  \d \Omega_5=\sin^2\theta\cos^2\theta \d \theta \d \hat{\bf x} \d
  \hat{\bf y}.
\label{eq:vol1}
\end{equation}
The HH have the explicit form
\begin{equation}
\Gamma_{KLM_L}^{l_xl_y} \left( \Omega_5 \right) = \psi_{K}^{l_xl_y}
( \theta ) Y_{LM_L}^{l_xl_y}(\hat{\bf x},\hat{\bf y}),
\label{eq:hh}
\end{equation}
where
\begin{equation}
Y_{LM_L}^{l_xl_y}(\hat{\bf x},\hat{\bf y})=
\left[ Y_{l_x} (\hat{\bf x}) \otimes Y_{l_y} (\hat{\bf y})
\right]_{LM_L},
\label{eq:y}
\end{equation}
and $Y_{lm}$ are spherical harmonics. In Eq.~\eqref{eq:hh} the
hyperangular functions are 
\begin{equation*} 
\psi_{K}^{l_xl_y} ( \theta ) = N_K^{l_xl_y}
\sin^{l_x} \theta \cos^{l_y} \theta
P_n^{l_x+1/2,l_y+1/2} (\cos 2\theta),
\end{equation*}
where $P_n^{\alpha,\beta} (x)$ are the Jacobi polynomials (see,
e.g.,~\cite{gra80}), $n = (K-l_x-l_y)/2$, and $N_K^{l_xl_y}$ are
normalization constants,
\begin{equation*}
N_K^{l_xl_y} = \sqrt{\frac{2(n!)(K+2)(n+l_x+l_y+1)!}
{\Gamma(n+l_x+3/2) \Gamma(n+l_y+3/2)}}.
\end{equation*}
The HH~\eqref{eq:hh} are
orthonormalized using the volume element~\eqref{eq:vol1}.
\subsection{E1 transition matrix elements}
The explicit expressions for the HH entering our ground state 
are the following (see~\eqref{eq:theta},~\eqref{eq:y} for notation):
\begin{equation}
\begin{split}
  \Gamma_{000}^{00}( \Omega_5 ) &= \frac{4}{\sqrt{\pi}} \: Y_{00}^{00}
  (\hat{\bf x},\hat{\bf y}), \\
  \Gamma_{200}^{00}( \Omega_5 ) &= \frac{8}{\sqrt{\pi}} \cos 2\theta
  \: Y_{00}^{00} (\hat{\bf x},\hat{\bf y}), \\ 
  \Gamma_{21M_L}^{11}( \Omega_5 ) &= \frac{8}{\sqrt{3\pi}} \sin 2\theta
  \: Y_{1M_L}^{11} (\hat{\bf x},\hat{\bf y}). 
\label{eq:ishh}
\end{split}
\end{equation}
The explicit expressions for the final state HH contributing to our
strength function are the following:
\begin{equation}
\begin{split}
  \Gamma_{11M_L}^{01}( \Omega_5 ) &= \frac{8}{\sqrt{2\pi}} \cos \theta
  \: Y_{1M_L}^{01} (\hat{\bf x},\hat{\bf y}), \\ 
  \Gamma_{11M_L}^{10}( \Omega_5 ) &= \frac{8}{\sqrt{2\pi}} \sin \theta
  \: Y_{1M_L}^{10} (\hat{\bf x},\hat{\bf y}), \\ 
  \Gamma_{31M_L}^{01}( \Omega_5 ) &= -\frac{8}{\sqrt{6\pi}} (4\cos
    2\theta + 1) \cos\theta \: Y_{1M_L}^{01} (\hat{\bf x},\hat{\bf y}),\\
  \Gamma_{31M_L}^{10}( \Omega_5 ) &= \frac{8}{\sqrt{6\pi}} (4\cos
    2\theta + 1) \sin\theta \: Y_{1M_L}^{10} (\hat{\bf x},\hat{\bf y}),\\
  \Gamma_{3LM_L}^{12}( \Omega_5 ) &= \frac{32}{\sqrt{6\pi}} \cos^2
    \theta \sin \theta \: Y_{LM_L}^{12} (\hat{\bf x},\hat{\bf y}),\qquad 
  L=1\,\,\mathrm{and}\,\, 2.
\label{eq:fshh}
\end{split}
\end{equation}
The E1 operator, Eq~\eqref{eq:reldipoleop}, can be written
as
\begin{equation*}
  \mathcal{M}(\uE 1, \mu) = - e Z_\uc \sqrt{\frac{2}{A (A-2)}} \rho \cos
  \theta Y_{1\mu} (\hat{\bf y}).
\end{equation*}
The ME 
\begin{equation}
  \langle K_f\,l_x\,l_{yf}\,L_f\,S,\,J_f=1,\,M_f=\mu|
  \cos \theta Y_{1\mu} (\hat{\bf y})|
  K_i\,l_x\,l_{yi}\,L_i\,S,\,J_i=0\rangle
  \label{angme}
\end{equation}
of the hyperangular part of this operator, between
basis states of the form
\begin{equation*}
  \left[ \Gamma_{KL}^{l_xl_y} \left( \Omega_5 \right) \otimes \theta_S
  \right]_{JM}
\end{equation*}
that include the HH~\eqref{eq:ishh} and~\eqref{eq:fshh} are required. We
denote these ME as
\begin{equation*}
  \langle K_f\,l_x\,l_{yf}\,L_f\,S | O |
  K_i\,l_x\,l_{yi}\,L_i\,S \rangle.
\end{equation*}
The non--zero ME are
\begin{equation}
  \begin{split}
    \langle 10110 | O | 00000 \rangle = \frac{1}{2\sqrt{2\pi}}, \qquad &
    \langle 10110 | O | 20000 \rangle = \frac{1}{4\sqrt{2\pi}}, \\
    \langle 30110 | O | 20000 \rangle = \frac{\sqrt{3}}{4\sqrt{2\pi}},
     \qquad &
    \langle 11011 | O | 21111 \rangle = \frac{1}{4\sqrt{2\pi}}, \\
    \langle 31011 | O | 21111 \rangle = \frac{1}{4\sqrt{6\pi}},  \qquad &
    \langle 31211 | O | 21111 \rangle = \frac{1}{4\sqrt{3\pi}}, \\
    \langle 31221 | O | 21111 \rangle = \frac{1}{4\sqrt{\pi}}.  \qquad &
  \end{split}
\label{eq:hme}
\end{equation}
The calculation with respect to spin variables and angles $\hat{\bf x}$,
$\hat{\bf y}$ is done using the formula (see, e.g.,~\cite{var89})
expressing the ME~\eqref{angme} in terms of reduced ME in the $\hat{\bf
x}, \hat{\bf y}$ subspace and then in terms of reduced ME in the
$\hat{\bf y}$ subspace. Finally, note that the results obtained in this
section are useful not only for \nuc{6}{He} but also when studying other
two--neutron halo nuclei.
}%
\bibliographystyle{phaip}
\bibliography{refs_he6}
%
\newpage
%
\begin{table}[htp]
  \begin{center}
    \caption{The model WFs for \nuc{6}{He}. The
    parameter $\kappa_0$ reproduces the two--neutron separation energy. 
    The last three columns give the hyperradius $\rho_\urms$, the matter
    radius $R_\urms$ and the distance
    between the core and the CM $r_\uc$.}
    \vspace{1ex}
    \begin{tabular}{ccccccccc}
      \hline\hline
      WF & $\kappa_0$ & $\kappa_1$ &
      $w_{00}$ & $w_{20}$ & $w_{21}$ & $\rho_\urms$ & $R_\urms$ &
      $r_\uc$ \\ 
      & (fm$^{-1}$) & (fm$^{-1}$) & & & & (fm) & (fm) & (fm) \\
      \hline
      $\Psi_1$ & 0.2163 & --- & 0 & 1 & 0 & 3.27 & 1.79 
      & 0.67 \\
      $\Psi_2$ & 0.2163 & 0.5370 & 0 & 1 & 0 & 5.37 
      & 2.50 & 1.10 \\
      $\Psi_3$ & 0.2163 & 0.5370 & 0.05 & 0.80 & 0.15 & 5.37 
      & 2.50 & 1.20 \\
      \hline\hline 
    \end{tabular}
  \label{tab:wfs}
  \end{center}
\vspace{-2.5mm}
\end{table}
\begin{table}[hbt]
\begin{center}
  \caption{The \nuc{6}{He} integrated strength 
  $\int_0^E\d E' \; \d B(\uE
  1)/\d E'$, where $E$ is the continuum energy, for
  different models and experimental data. 
  The last column corresponds to $E=\infty$ and equals to the result of
  Eq.~\eqref{eq:newcsr}. All
  values are given in $e^{2}\mathrm{fm}^2$.}
  \vspace{1ex}
  \begin{minipage}{0.8\textwidth}
  \begin{tabular}{l| r@{.}l r@{.}l c}
    \hline
    \hline
    & \multicolumn{2}{c}{$E=4$~MeV} & \multicolumn{2}{c}{$E=9$~MeV} &
    Total B(E1) \\
    \hline
    Aumann {\textit et al}~\cite{aum99:59} & \hspace{3mm}0&59(12)  &
    \hspace{4mm}1&2(2) \\ 
    $\Psi_{1}$ & 0&079 & 0&18 & 0.43 \\
    $\Psi_{2}$ & 0&29 & 0&63 & 1.14 \\
    $\Psi_{3}$ & 0&42 & 0&83 & 1.38 \\
    Realistic WF\footnote{Danilin \textit{et al.}~\cite{dan98:632}} &
    0&71 & 1&02 & 1.37 \\
    \hline
    \hline
  \end{tabular}
  \end{minipage}
  \label{tab:he6strength}
\end{center}
\vspace{-2.5mm}
\end{table}
\begin{table}[hbt]
\begin{center}
  \caption{The \nuc{6}{He} EMD cross section obtained with a lead target at
  different energies. The theoretical values are obtained by
  integrating~\eqref{eq:xsec} up to 10~MeV.}
  \vspace{1ex}
  \begin{minipage}{0.8\textwidth}
\begin{tabular}{c | c c | c}
    \hline
    \hline
    Energy (MeV/A) & \multicolumn{2}{|c|}{Exp.} & This work \\
    & $\sigma_\mathrm{EMD}$ (mb) & Ref. & $\sigma_\mathrm{EMD}$ (mb) \\
    \hline
    37 & 830\footnote{80\% of the incident \nuc{6}{He} ions were in the range
    28--52 MeV/A. A Glauber--type calculation estimated that 60\% of the
    total inelastic cross section were due to EMD.}
         &~\cite{war00:62} & 551 \\
    240 & $520 \pm 110$\footnote{The EMD cross section was obtained by
    subtracting eikonal model cross section~\cite{ber98:57} (for nuclear
    excitations) from the measured inelastic excitation
    cross sections.}
        &~\cite{aum99:59} & 333 \\
    \hline
    \hline
  \end{tabular}
  \end{minipage}
  \label{tab:he6xsec}
\end{center}
\vspace{-2.5mm}
\end{table}
%
\end{document}